\documentclass[]{tPHM2e}

\usepackage[usenames,dvipsnames]{color}
\usepackage{amsmath}
\usepackage{float}
\usepackage{url}

\newcommand{\bk}{\mathbf{k}}
\newcommand{\ua}{\uparrow}
\newcommand{\da}{\downarrow}
\newcommand{\eq}{\begin{equation}}
\newcommand{\eqx}{\end{equation}}
\newcommand{\eqn}{\begin{eqnarray}}
\newcommand{\eqnx}{\end{eqnarray}}
\newcommand{\s}{\sigma}

\newcommand{\veck}{{\bf k}}
\newcommand{\veci}{{\bf i}}
\newcommand{\vecj}{{\bf j}}

\newcommand{\vecm}{{\bf m}}
\newcommand{\vecl}{{\bf l}}
\newcommand{\ra}{\rangle}
\newcommand{\la}{\langle}

\newcommand{\dg}{\dagger}

\newcommand{\Ps}{|\Psi_0\rangle}
\newcommand{\Heff}{\hat{H}_0^{\rm eff}}

\begin{document}

\title{Comparison of two approaches for the treatment of Gutzwiller variational wave functions}

\author{J.~Kaczmarczyk$^{\rm a}$$^{\ast}$,
\thanks{$^\ast$ Email: jan.kaczmarczyk@uj.edu.pl\vspace{6pt}}\\
$^{\rm a}$ Instytut Fizyki im. Mariana Smoluchowskiego, Uniwersytet Jagiello\'{n}ski, Reymonta 4, 30-059 Krak\'ow, Poland
}
\maketitle

\begin{abstract}
In this work we analyze the variational problem emerging from the Gutzwiller approach to strongly correlated systems. This problem comprises the two main steps: {\it evaluation} and {\it minimization} of the ground state energy $W$ for the postulated Gutzwiller Wave Function (GWF). We discuss the available methods for evaluating $W$, in particular the recently proposed diagrammatic expansion method. We compare the two existing approaches to minimize $W$: the standard approach based on the effective single-particle Hamiltonian (EH) and the so-called Statistically-consistent Gutzwiller Approximation (SGA). On the example of the superconducting phase analysis we show that these approaches lead to the same minimum as it should be. However, the calculations within the SGA method are easier to perform and the two approaches allow for a simple cross-check of the obtained results. Finally, we show two ways of solving the equations resulting from the variational procedure, as well as how to incorporate the condition for a fixed number of particles.
\end{abstract}

\section{Introduction}
Systems with strong electron correlations are in the center of interest of condensed matter physicists for many decades now. The milestones of this field were the discoveries of heavy-fermion systems \cite{Ott} and high-temperature superconductors \cite{Bednorz}. Because of the great complexity of those systems, there is up to date no universal theoretical approach to describe them. Among the methods developed by theoreticians are the Dynamical Mean Field Theory (DMFT) \cite{RevModPhys.68.13}, Gutzwiller approach \cite{Gutzwiller,PhysRevLett.86.2605,TDGAmulti,TDGA2}, and variational Monte Carlo (VMC) methods \cite{EdeggerRev}, as well as the combinations of DMFT and Gutwiller methods with Density Functional Theory (DFT) \cite{RevModPhys.78.865,PhysRevB.77.073101}. Both the Gutzwiller approach and VMC are variational approaches, in which the central tasks are the postulation of the variational wave function $| \Psi \ra$, as well as the \textit{evaluation} and \textit{minimization} of the ground state energy, $W \equiv \la \Psi| \hat{H} | \Psi \ra$.

In this paper, the minimization task is analyzed in detail. First, we introduce the Gutzwiller method for the Hubbard model. Second, we discuss selected approaches to evaluate $W$. We concentrate on one of them proposed very recently, the Diagrammatic Expansion for Gutzwiller Wave Function (DE-GWF) \cite{Buenemann, PhysRevB.88.115127, DEtJ}. Third, we discuss two approaches to minimizing $W$, which is a complicated object containing long-range real-space correlations. On the example of the superconducting phase analysis we show equivalence of these two approaches: they lead to the same equations (minimization conditions). Finally, we discuss two schemes of solving these equations differing in complexity and applicability, as well as we show how to efficiently incorporate the condition for a fixed number of particles into one of them.

\section{Gutzwiller Wave Function}

The systems with strong electron correlations are very often described starting from the single-band Hubbard Hamiltonian, which for the high-temperature superconductors is an effective model of the Cu-O$_2$ plane
\eq
\hat{H}=\hat{H}_0 + U\sum_{\veci} \hat{d}_{\veci} \,,
\hat{H}_0=\sum_{\veci,\vecj,\sigma}t_{\veci,\vecj}\hat{c}_{\veci,\sigma}^{\dagger}
\hat{c}_{\vecj,\sigma}^{\phantom{\dagger}} \, ,
\hat{d}_{\veci}\equiv\hat{n}_{\veci,\uparrow}\hat{n}_{\veci,\downarrow} \, ,
\label{eq:1}
\end{equation}
where $\veci = (i_1, i_2)$ is the two-dimensional site-index and $\s = \ua, \da$ is the spin quantum number. Typically the Coulomb interaction $U$ is much larger than the hopping amplitude between the nearest neighbours. In such a situation it is favorable energetically to decrease the weight of configurations with doubly-occupied sites in the wave function of the system. This can be achieved by using the Gutzwiller Wave Function (GWF)~\cite{Gutzwiller}, which has the form
\begin{equation}
|\Psi_{\rm G}\rangle = \hat{P} |\Psi_0\rangle =
\prod\nolimits_{\veci}\hat{P}_{\veci}|\Psi_0\rangle = \prod\nolimits_{\veci} \left[ 1- \left( 1 - g \right) \hat{d}_{\veci} \right]|\Psi_0\rangle\; ,
\label{eq:1.2}
\end{equation}
where $g$ is a variational parameter and $|\Psi_0\rangle$ is a single-particle product state (``Slater determinant'') to be defined later. However, an alternative choice was shown to lead to much faster convergence \cite{Buenemann}. Namely, we define a local Gutzwiller correlator $\hat{P}_\veci$ in such a way that it obeys the following equation \cite{Gebhard}
\eq
\hat{P}^2_{\veci} \equiv 1+x\hat{d}_{\veci}^{\rm HF} \; , \label{eq:proj2}
\eqx
where $x$ is a variational parameter and the Hartree--Fock (HF) operators are defined by $\hat{d}_{\veci}^{\rm HF} \equiv \hat{n}^{\rm HF}_{\veci,\uparrow}\hat{n}^{\rm HF}_{\veci,\downarrow}$ and $\hat{n}^{\rm HF}_{\veci,\sigma}\equiv\hat{n}_{\veci,\sigma}-n_0 $ with $n_0 \equiv \langle \hat{n}_{\veci,\sigma} \rangle_0 \equiv \langle \Psi_0| \hat{n}_{\veci,\sigma} |\Psi_0 \rangle$.

Within the variational approach with GWF the principal task is the evaluation of the ground state energy
\eq
W \equiv \la \hat{H} \ra_{\rm G} \equiv \frac{\la \Psi_{\rm G} | \hat{H} |\Psi_{\rm G} \ra}{\la \Psi_{\rm G} | \Psi_{\rm G} \ra} \equiv \frac{ \la \Psi_{0} | \hat{P} \hat{H} \hat{P} |\Psi_{0} \ra }{ \la \Psi_{0} | \hat{P}^2 |\Psi_{0} \ra } \label{eq:W}.
\eqx
The most important difference among methods based on GWF is the approach taken towards computing $W$.
In the Gutzwiller approximation (GA) average values of the Hamiltonian terms are approximated by a product of some function (called for the hopping term the \textit{Gutzwiller band narrowing factor}, $g_t$) and their average values in the non-correlated wave function $| \Psi_0 \rangle$, e.g. $\la \hat{c}^\dg_{\veci\s} \hat{c}_{\vecj\s} \ra_G \approx g_t \la \hat{c}^\dg_{\veci\s} \hat{c}_{\vecj\s} \ra_0$. This yields a very fast method, but is also the reason behind the inability to describe within GA the superconducting phase in the Hubbard model \cite{PhysRevB.88.115127}.

In the VMC method, $W$ is calculated with the Monte Carlo sampling technique. Such an approach is much more accurate than GA, but suffers from the following drawbacks: (i) low speed (as compared to GA); (ii) difficulty (or inability) to perform calculations for nonzero temperature; (iii) necessity of using finite-size lattices, which can lead to large errors, e.g. for the states with Fermi surface deformations (the so-called Pomeranchuk phase) \cite{PhysRevB.74.165109}. The advantage of VMC is the possibility to include additional Jastrow factors in the trial wave function \cite{Yokoyama2,Baeriswyl,Hetenyi}.

In the recently proposed \cite{Buenemann,PhysRevB.88.115127,DEtJ} DE-GWF technique, an alternative approach is taken, which can remedy all the above problems. Namely, with the use of the correlator defined by (\ref{eq:proj2}) and after applying the linked-cluster theorem \cite{Fetter}, we are able \cite{Buenemann} to calculate the ground state energy $W(|\Psi_0\rangle,x) \equiv \langle \hat{H} \rangle_{\rm G}$ and the generalized grand potential $\mathcal{F}$ at zero temperature for a translationally invariant system
\eqn
\langle \hat{H} \rangle_{\rm G}
&=& 2\sum_{\veci,\vecj}t_{\veci,\vecj}
\left[q^2T_{\veci,\vecj}^{(1),(1)}
+q\alpha T_{\veci,\vecj}^{(1),(3)}
+q\alpha T_{\veci,\vecj}^{(3),(1)}
+\alpha^2 T_{\veci,\vecj}^{(3),(3)}\right]
\nonumber \\ &&
+LU \lambda_d^2\left[(1 - x d_0)I^{(4)}_\veci+2n_0I^{(2)}_\veci+d_0\right] \; , \\
\mathcal{F} &=& \langle \hat{H} \rangle_{\rm G} - 2 \mu_G n_G L\;, \label{eq:erz2}
\eqnx
with
\eqn
n_G &\equiv& \langle \hat{n}_{\veci,\sigma} \rangle_{\rm G}=
\lambda^2_{d}\left[d_0+I^{(4)}_\veci(1-xd_0)+2n_0I^{(2)}_\veci\right]
\nonumber \\ &&
+\lambda^2_{1}
\left[m^0_{1}+I^{(2)}_\veci(1-2n_0)-I^{(4)}_\veci(1+x m^0_{1})\right]\;,
\eqnx
where $L$ is the number of sites, $n_G$ is the average number of particles in the correlated state per site
\footnote{Note that for superconducting states the correlated and non-correlated numbers of particles ($n_G$ and $n_0$) may differ, and hence we minimize the functional $\mathcal{F}$ at a constant chemical potential $\mu_G$, and not the energy $W$ at a constant number of particles $n_G$.},
$m^0_{1} = n_0(1-n_0)$, $d_0 \equiv n_0^2$, and
$q$, $\alpha$, $\lambda_1$, and $\lambda_d$ are functions of $n_0$ and $x$ \cite{Buenemann}.
The diagrammatic sums (DS) appearing in the above expressions are defined by
\begin{equation}
S_{\veci[,\vecj]}=\sum_{k=0}^{\infty}\frac{x^k}{k!}S_{\veci[,\vecj]}(k), \label{eq:expansion}
\end{equation}
where
\begin{equation}
S_{\veci[,\vecj]}\in\{I_\veci^{(2)}, I_\veci^{(4)}, T_{\veci, \vecj}^{(1),(1)}, T_{\veci, \vecj}^{(1),(3)}, T_{\veci, \vecj}^{(3),(1)}, T_{\veci, \vecj}^{(3),(3)}
\}.
\end{equation}
The $k$-th order sum contributions have the following forms:
{\arraycolsep=2pt\begin{eqnarray}\label{ds1}
I_\veci^{(2)[(4)]}(k)&\equiv&\sum_{\vecl_1,\ldots, \vecl_k}
\bigl\langle
\hat{n}^{\rm HF}_{\veci,\sigma}[\hat{n}^{\rm HF}_{\veci, \overline{\sigma}}]
\hat{d}^{\rm HF}_{\vecl_1,\ldots,\vecl_k}
\bigr\rangle^{\rm c}_{0}\; ,\\
\label{ds2}
T_{\veci,\vecj}^{(1)[(3)],(1)[(3)]}(k)&\equiv&
\sum_{\vecl_1,\ldots,\vecl_k}
\bigl\langle
[\hat{n}^{\rm HF}_{\veci,\bar{\sigma}}]
\hat{c}^{\dagger}_{\veci,\sigma}
[\hat{n}^{\rm HF}_{\vecj,\bar{\sigma}}]
\hat{c}_{\vecj,\sigma}^{\phantom{\dagger}}\hat{d}^{\rm HF}_{\vecl_1,\ldots,\vecl_k}
\rangle^{\rm c}_{0}\;,
\end{eqnarray}}%
where the notation $(1)[(3)]$ means that when calculating for the index (3) also the term in square brackets needs to be taken, e.g. $T_{\veci,\vecj}^{(1),(3)}(k) \equiv
\sum_{\vecl_1,\ldots,\vecl_k}
\bigl\langle
\hat{c}^{\dagger}_{\veci,\sigma}
\hat{n}^{\rm HF}_{\vecj,\bar{\sigma}}
\hat{c}_{\vecj,\sigma}^{\phantom{\dagger}}\hat{d}^{\rm HF}_{\vecl_1,\ldots,\vecl_k}
\rangle^{\rm c}_{0}$.
To calculate the DS from (\ref{ds1})-(\ref{ds2}) we apply the Wick's theorem \cite{Fetter} and perform the summation over $\vecl_1$, ..., $\vecl_k$ on a square lattice \cite{DEtJ}. The $k$-th order terms of (\ref{eq:expansion}) correspond to diagrams with one (or two) external vertices on sites $\veci$ (or $\veci$ and $\vecj$) and $k$ internal vertices. The notation $\langle \dots \rangle_0^{ \rm c}$ in (\ref{ds1})-(\ref{ds2}) indicates that only the connected diagrams are to be kept.
The vertices of diagrams are connected with lines (corresponding to contractions from Wick's theorem), which in the case of the superconducting state with intersite pairing are given by
\eq    \label{pl}
P_{\vecl,\vecl'} \equiv P^{\sigma}_{\vecl,\vecl'}  \equiv
\langle\hat{c}^{\dagger}_{\vecl,\sigma}\hat{c}_{\vecl',\sigma}^{\phantom{\dagger}}\rangle_0
-\delta_{\vecl,\vecl'}n_0\;, \qquad
S_{\vecl,\vecl'}  \equiv
\langle\hat{c}^{\dagger}_{\vecl,\ua}\hat{c}^{\dagger}_{\vecl',\da}\rangle_0.
\eqx
%
Due to translational invariance of the system we only need to calculate DS with fixed $\veci$ and $\vecj$. In other words, $I_\veci^{(2)[(4)]} \equiv I^{(2)[(4)]}$ and $T_{\veci,\vecj}^{(1)[(3)],(1)[(3)]} \equiv T_{\veci-\vecj}^{(1)[(3)],(1)[(3)]} $.
To be able to calculate $W$ in a particular physical situation, we need to introduce two additional approximations. First, the summation in (\ref{eq:expansion}) has to be performed up to a certain order $k$. Second, the lines (\ref{pl}) have to be included up to a certain cutoff distance. Consequently, the summations in (\ref{eq:expansion}) and (\ref{ds1})-(\ref{ds2}) become finite.

As a result of our diagrammatic expansion procedure, $W$ (or $\mathcal{F}$) is a function of the variational parameter $x$ and the non-correlated wave function $\Ps$. This wave function enters into the variational problem via $n \equiv \la \hat{n}_{i\s} \ra_0$ and the lines $P_{\vecl,\vecl'}$ and $S_{\vecl,\vecl'}$.
We now come to the central question of the paper: how to minimize the functional $\mathcal{F}$? We discuss here two minimization procedures, which are equivalent in the sense that they lead to the same minimum (as will be shown below). On the other hand, they differ in important technical aspects, one of them is easier to perform, and they can be used to make a simple cross-check of the obtained results.
The minimization task is also present in various forms of the GA approaches \cite{Fukushima,Ogata2} (also known by the name of Renormalized Mean-Field Theory, RMFT \cite{PhysRevB.37.3759,EdeggerRev}), and therefore the conclusions of this paper hold for them as well.
The minimization over $x$ is a one-dimensional minimization, which yields the $x_0$ value of that parameter ($x_0$ depends on $\Ps$, so we need to minimize with respect to $x$ every time $\Ps$ is changed). In the following we consider $\mathcal{F}(x_0, \Ps)$, which is to be miminized with respect to $\Ps$.

\subsection{Effective Hamiltonian (EH) approach}

We start with the approach used commonly in the literature (cf. e.g. \cite{Yang, Wang, PhysRevB.67.075103, negativeU, HeffAlternative, Buenemann, PhysRevB.88.115127, DEtJ}), which we will call the \textit{EH scheme}. The condition for the minimum with respect to $\Ps$ is the following
\eq
\frac{\delta\left[ \mathcal{F} - \lambda \left( \la \Psi_0 | \Psi_0 \ra - 1 \right) \right] }{\delta \la \Psi_0 |} = 0. \label{eq:min}
\eqx
Interpreting $\mathcal{F}$ as a composed function of the lines $P_{\vecl,\vecm}$, $S_{\vecl,\vecm}$ and $\Ps$, and using for the derivatives of the lines e.g. $\delta S_{\vecl,\vecm} / \delta \la \Psi_0 | = \delta ( \la \Psi_0 | \hat{c}^{\dagger}_{\vecl,\ua}\hat{c}^{\dagger}_{\vecm,\da} | \Psi_0 \ra ) / \delta \la \Psi_0 | = \hat{c}^{\dagger}_{\vecl,\ua}\hat{c}^{\dagger}_{\vecm,\da} | \Psi_0 \ra$ we obtain from (\ref{eq:min}) the following equation\footnote{Note that there exists an alternative derivation of the effective Hamiltonian, see e.g. \cite{HeffAlternative}, Appendix C.}
\eq
\sum_{\veci,\vecj, \s} \frac{\partial \mathcal{F}}{\partial P_{\veci,\vecj}}
\hat{c}^{\dagger}_{\veci,\s}\hat{c}_{\vecj,\s} | \Psi_0 \ra
 + \sum_{\veci,\vecj} \left( \frac{\partial \mathcal{F}}{\partial S_{\veci,\vecj}}
 \hat{c}^{\dagger}_{\veci,\ua}\hat{c}^{\dagger}_{\vecj,\da} + {\rm H.c.}\right) | \Psi_0 \ra = \lambda \Ps.
\eqx
It is the effective single-particle Schr\"{o}dinger equation
\eq \label{eq:iou}
\hat{H}_0^{\rm eff} |\Psi_0\rangle= E^{\rm eff} |\Psi_0\rangle,
\eqx
with the self-consistently defined effective single-particle Hamiltonian
\begin{eqnarray}
\hat{H}_0^{\rm eff} &=& \sum_{\veci,\vecj, \s}t^{\rm eff}_{\veci,\vecj}\label{eq:iou0}
\hat{c}_{\veci,\sigma}^{\dagger}\hat{c}_{\vecj,\sigma}^{\phantom{\dagger}} + \sum_{\veci,\vecj} \big( \Delta^{\rm eff}_{\veci,\vecj} \hat{c}_{\veci,\uparrow}^{\dagger}\hat{c}_{\vecj,\downarrow}^{\dagger} + {\rm H.c.} \big),\\
t^{\rm eff}_{\veci,\vecj} &=& \frac{\partial \mathcal{F} }{\partial P_{\veci,\vecj}} \;, \qquad
\Delta^{\rm eff}_{\veci,\vecj} = \frac{\partial \mathcal{F} }{\partial S_{\veci,\vecj}} \;.  \label{eq:iouF}
\end{eqnarray}
As can be seen from (\ref{eq:iou}), $\Ps$ is the ground state of $\Heff$. After Fourier transformation and in the Nambu representation the Hamiltonian takes the form
\eq
\hat{H}_0^{\rm eff} = \sum_{\bk} ( \hat{c}_{\bk \ua}^\dg, \hat{c}_{-\bk \da} )
\left( \begin{array}{cc}
\epsilon_\bk & \Delta_\veck \\
\Delta_\veck^* & -\epsilon_\bk
\end{array} \right)
\left( \begin{array}{c} \hat{c}_{\bk \ua} \\ \hat{c}_{-\bk \da}^\dg \end{array} \right)
+ \sum_\bk \epsilon_\bk, \label{eq:HeffNambu}
\eqx
where the effective dispersion relation and the effective gap are defined as
\begin{eqnarray}
\epsilon_\bk &=& \frac{1}{L}\sum_{\veci,\vecj}\exp^{{\rm i}(\veci-\vecj)\veck} t^{\rm eff}_{\veci,\vecj} = \left[ \sum_{\vecj}\exp^{{\rm i}(\veci-\vecj)\veck} t^{\rm eff}_{\veci,\vecj} \right]_{\veci = (0,0)}, \label{eq:effdispersion}
\\
\label{eq:effgap}
\Delta_\veck &=& \frac{1}{L}\sum_{\veci,\vecj}\exp^{{\rm i}(\veci-\vecj)\veck} \Delta^{\rm eff}_{\veci,\vecj} = \left[ \sum_{\vecj}\exp^{{\rm i}(\veci-\vecj)\veck} \Delta^{\rm eff}_{\veci,\vecj} \right]_{\veci = (0,0)}.
\end{eqnarray}
The last expressions for $\epsilon_\bk$ and $\Delta_\veck$ are valid for a homogeneous system. Diagonalization leads to the eigenvalues $E_{\bk, i} \in \{E_\bk, - E_\bk\}$, where the excitation energies $E_\veck$ are given by
\eq
E_\veck = \sqrt{\epsilon_\bk^2 + \Delta_\veck^2}. \label{eq:exc}
\eqx
To calculate the new $|\Psi_0\rangle$ lines [from their definition in Eq. (\ref{pl})] we need to know the Bogolyubov-de Gennes (BdG) transformation coefficients (in contrast to the alternative minimization procedure described in Section 2.2)
\eq
\left( \begin{array}{c} \hat{c}_{\bk \ua} \\ \hat{c}_{-\bk \da}^\dg \end{array} \right) =
\left( \begin{array}{cc}
u_\bk & -v_\veck \\
v_\bk & u_\bk
\end{array} \right)
\left( \begin{array}{c} \hat{\alpha}_{\bk} \\ \hat{\beta}_{\bk}^\dg \end{array} \right), \label{eq:BdG}
\eqx
where $\hat{\alpha}_{\bk}$ and $\hat{\beta}_{\bk}$ are new quasiparticle operators, and the transformation coefficients are given by $u_\bk = \frac{1}{\sqrt{2}} \left( 1 + \epsilon_\bk / E_\veck \right)^{1/2}$, $v_\bk = \frac{1}{\sqrt{2}} \left( 1 - \epsilon_\bk / E_\veck \right)^{1/2}$.
Using the definition (\ref{pl}) together with (\ref{eq:BdG}) leads to the prescriptions
\eqn
P_{\vecl, \vecm} &=& \frac{1}{L} \sum_\bk e^{i \bk (\vecl - \vecm)} n_{\bk}^0, \qquad
n_{\bk}^0 = \frac{1}{2} \Big( 1 - \frac{\epsilon_\bk}{E_\veck} \Big),
\label{eq:PLine} \\
S_{\vecl, \vecm} &=& \frac{1}{L} \sum_\bk e^{i \bk (\vecl - \vecm)} \Delta_{\bk}^0, \qquad
\Delta_{\bk}^0 = - \frac{1}{2} \frac{\Delta_\veck}{E_\veck}
\label{eq:SLine}.
\eqnx
These equations are to be solved together with (\ref{eq:iouF}) and the solution procedures will be discussed in Section 3.

Please note that the above minimization procedure has been used commonly in the literature \cite{Yang, Wang, PhysRevB.67.075103, negativeU}. The differences between DE-GWF and those approaches lie in the method of calculating $W$ and the number of $\Ps$ lines included. Within DE-GWF, due to the way of calculating $W$, many $\Ps$ lines are included (e.g. up to seventh neighbors in \cite{PhysRevB.88.115127} or fourteenth neighbors in \cite{DEtJ}).

\subsection{Method based on the Lagrange multipliers}

The alternative derivation of the minimization conditions is based on the so-called Statistically-consistent Gutzwiller Approximation (SGA) \cite{PhysRevB.84.125140,Wysokinski,Abram,Zegrodnik,Zegrodnik2,Zegrodnik3,Kadzielawa,Olga,Olga2,WysokinskiVietri,Jedrak2,Jedrak1,SGA,preSGA}, and therefore we call it the \textit{SGA scheme}. Within this method we also start with the expression for $W$ (or $\mathcal{F}$) and supply it with the Lagrange-multiplier terms yielding the following auxiliary energy operator
\eq
\hat{K} = \mathcal{F} +
\sum_{\veci,\vecj, \s}
t^{\rm eff}_{\veci,\vecj}
\left( \hat{c}_{\veci,\sigma}^{\dagger}\hat{c}_{\vecj,\sigma} - P_{\veci, \vecj} \right) +
\sum_{\veci,\vecj} \left[
\Delta^{\rm eff}_{\veci,\vecj}
\left( \hat{c}_{\veci,\uparrow}^{\dagger}\hat{c}_{\vecj,\downarrow}^{\dagger} - S_{\veci, \vecj} \right)
+ {\rm H.c.}
\right].
\eqx
In this formulation the effective parameters $t^{\rm eff}_{\veci,\vecj}$ and $\Delta^{\rm eff}_{\veci,\vecj}$ play the role of Lagrange multipliers ensuring that the average values of the operators (e.g. $\hat{c}_{\veci,\sigma}^{\dagger}\hat{c}_{\vecj,\sigma}$) within the wave function $\Ps$ are equal to the lines (e.g. $P_{\veci, \vecj}$). For formal details of such procedure see \cite{JJPHD}. After Fourier transformation and in the Nambu representation $\hat{K}$ has the form
\eqn
\hat{K} &=& \mathcal{F} + \sum_{\bk} ( \hat{c}_{\bk \ua}^\dg, \hat{c}_{-\bk \da} )
\left( \begin{array}{cc}
\epsilon_{\bk} & \Delta_\bk \\
\Delta^*_\bk & -\epsilon_\bk
\end{array} \right)
\left( \begin{array}{c} \hat{c}_{\bk \ua} \\ \hat{c}_{-\bk \da}^\dg \end{array} \right)
+ \sum_\bk \epsilon_\bk \\
&& - \sum_{\veci, \vecj, \s} t^{\rm eff}_{\veci,\vecj} P_{\veci,\vecj} -
\sum_{\veci, \vecj} \left( \Delta^{\rm eff}_{\veci,\vecj} S_{\veci,\vecj} + {\rm H.c.}\right), \nonumber
\eqnx
with $\epsilon_{\bk}$ and $\Delta_\bk$ as in (\ref{eq:effdispersion}) and (\ref{eq:effgap}).
The diagonalization procedure gives the eigenenergies $E_{\bk, i} \in \{E_\bk, -E_\bk\}$, with $E_\bk$ as in (\ref{eq:exc}). The generalized grand potential functional $\mathfrak{F}$ for operator $\hat{K}$ is as follows
\eqn
\mathfrak{F} &=& f_\beta + \mathcal{F} + \sum_{\bk} \epsilon_\bk - \sum_{\veci, \vecj, \s} t^{\rm eff}_{\veci,\vecj} P_{\veci,\vecj} -
\sum_{\veci, \vecj} \left( \Delta^{\rm eff}_{\veci,\vecj} S_{\veci,\vecj} + {\rm H.c.}\right), \\
f_\beta &=& -\beta^{-1} \sum_{\bk, i=1,2} \ln{\left(1+e^{-\beta E_{\bk, i}}\right)},
\eqnx
where $\beta = 1 / (k_B T)$ is the inverse temperature\footnote{The nonzero temperature is introduced for technical reasons. In the following we will take the zero-temperature limit.}. This functional is minimized with respect to the lines $P_{\veci,\vecj}$ and $S_{\veci,\vecj}$, as well as with respect to the Lagrange multipliers $t^{\rm eff}_{\veci,\vecj}$ and $\Delta^{\rm eff}_{\veci,\vecj}$:
\eq
\frac{\partial \mathfrak{F}}{\partial P_{\veci,\vecj}} = 0, \quad
\frac{\partial \mathfrak{F}}{\partial S_{\veci,\vecj}} = 0, \quad
\frac{\partial \mathfrak{F}}{\partial t^{\rm eff}_{\veci,\vecj}} = 0, \quad
\frac{\partial \mathfrak{F}}{\partial \Delta^{\rm eff}_{\veci,\vecj}} = 0.
\eqx
These conditions yield, respectively,
\eqn
t^{\rm eff}_{\veci,\vecj} &=& \frac{\partial \mathcal{F}}{\partial P_{\veci,\vecj}}, \qquad \qquad \quad \,\,
\Delta^{\rm eff}_{\veci,\vecj} = \frac{\partial \mathcal{F}}{\partial S_{\veci,\vecj}}, \label{eq:minCond1}
\\
P_{\veci,\vecj} &=& \frac{\partial f_\beta}{\partial t^{\rm eff}_{\veci,\vecj}}
+ \sum_{\bk} \frac{\partial \epsilon_\bk}{\partial t^{\rm eff}_{\veci,\vecj}} , \quad
S_{\veci,\vecj} = \frac{\partial f_\beta}{\partial \Delta^{\rm eff}_{\veci,\vecj}}. \label{eq:minCond2}
\eqnx
It is clear that the equations (\ref{eq:minCond1}) are exactly the same as in (\ref{eq:iouF}). The remaining two can be expressed as follows
\eqn
P_{\veci,\vecj} &=& \sum_\bk \left[ \sum_{i=1,2} f(E_{\bk, i}) \frac{\partial E_{\bk, i}}{\partial t^{\rm eff}_{\veci,\vecj}} + \frac{\partial \epsilon_{\bk}}{\partial t^{\rm eff}_{\veci,\vecj}} \right], \\
S_{\veci,\vecj} &=& \sum_{\bk, i=1,2} f(E_{\bk, i}) \frac{\partial E_{\bk, i}}{\partial \Delta^{\rm eff}_{\veci,\vecj}},
\eqnx
where $f(E) = 1/\left[1+e^{(\beta E)}\right]$ is the Fermi-Dirac distribution function. Evaluating the derivatives, we obtain
\eqn
P_{\veci,\vecj} &=& \sum_\bk\left\{ \frac{\epsilon_\bk}{2 E_\bk} \left[ 2 f(E_\bk) - 1 \right] + \frac{1}{2} \right\} \cos{\left[ (\veci - \vecj) \bk \right]}, \\
S_{\veci,\vecj} &=& \sum_\bk\left\{ \frac{\Delta_\bk}{2 E_\bk} \left[ 2 f(E_\bk) - 1 \right]  \right\} \cos{\left[ (\veci - \vecj) \bk \right]}.
\eqnx
If the zero-temperature limit is taken ($\beta \to \infty$), the $f(E_\bk)$ terms vanish and the above equations become
\eqn
P_{\veci,\vecj} &=& \sum_\bk\left[ \frac{1}{2}\left(1 - \frac{\epsilon_\bk}{E_\bk}\right) \right] \cos{\left[ (\veci - \vecj) \bk \right]}, \label{eq:PL_SGA}\\
S_{\veci,\vecj} &=& \sum_\bk\left( - \frac{1}{2} \frac{\Delta_\bk}{E_\bk}\right) \cos{\left[ (\veci - \vecj) \bk \right]}.\label{eq:SL_SGA}
\eqnx
They are the same as (\ref{eq:PLine})-(\ref{eq:SLine}). Therefore, in the end, both the EH and SGA schemes yield the same equations and are equivalent. However, there are significant differences in deriving these equations. First, in the EH scheme we need to perform the BdG transformation, for which we need both the eigenvalues and eigenvectors of the matrix in the effective Hamiltonian (\ref{eq:HeffNambu}). This is technically more complicated and error-prone, than just finding the eigenvalues, and their derivatives over $t^{\rm eff}_{\veci,\vecj}$ and $\Delta^{\rm eff}_{\veci,\vecj}$ as in the SGA scheme. Second, in some situations\footnote{For example, in a system with coexistence of superconductivity and magnetism \cite{PhysRevB.84.125140,Olga} with both ferro- and antiferromagnetic order.}
the Hamiltonian matrix contains so many independent variables that the analytical calculation of the eigenenergies and BdG coefficients is inpractical (or impossible). In such cases in the EH scheme a numerical diagonalization (with finding the eigenvectors) has to be performed to find the expressions for new lines. This is again more complicated than the SGA scheme alternative, which is evaluating the derivatives of $f_\beta$ over $t^{\rm eff}_{\veci,\vecj}$ and $\Delta^{\rm eff}_{\veci,\vecj}$ in (\ref{eq:minCond2}) numerically (for this task only finding the eigenvalues is necessary). However, in this case analytical computation of the derivatives of $\mathcal{F}$ over the lines in (\ref{eq:minCond1}) is necessary to avoid too much precision loss from computing derivatives numerically twice.

In any case, two distinct methods of obtaining the minimization conditions allow to perform reliable cross-check of the results. A quick test of the EH approach would be to verify the lines obtained from the definition [here (\ref{eq:PLine})-(\ref{eq:SLine})] with those obtained via numerical differentiation in (\ref{eq:minCond2}).

\section{Solving the minimization conditions}

\begin{figure}[ht!]
\centering
\includegraphics[width=0.55\columnwidth]{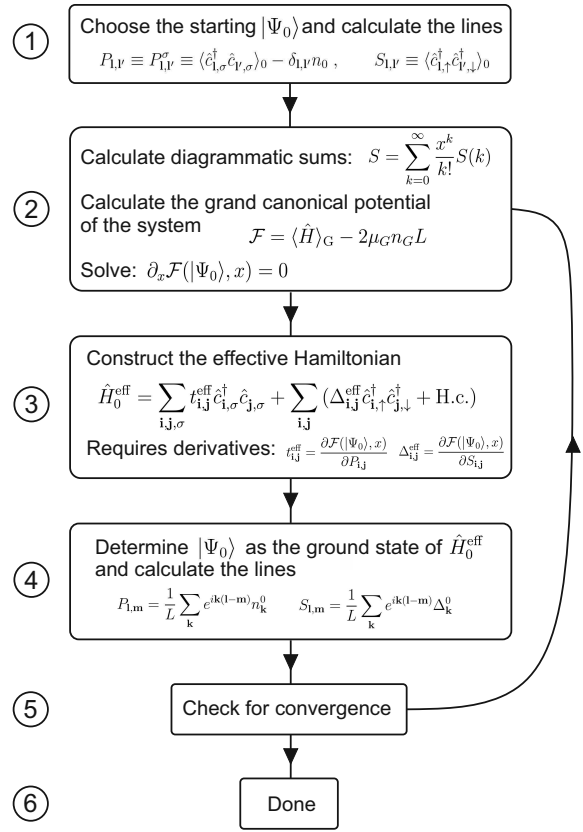}
\caption{The self-consistency loop of the DE-GWF method.}
\label{fig:3}
\end{figure}

The equations (\ref{eq:minCond1}) and (\ref{eq:PL_SGA})-(\ref{eq:SL_SGA}), or equivalently (\ref{eq:iouF}) and (\ref{eq:PLine})-(\ref{eq:SLine}), are to be solved to study properties of the investigated phase. The solution scheme used so far \cite{Buenemann,PhysRevB.88.115127,DEtJ} implements the following self-consistent procedure: (1) We start with $\Ps$ being the ground state of some fictitious effective Hamiltonian, and calculate the lines of such wave function from (\ref{eq:PL_SGA})-(\ref{eq:SL_SGA}); (2) For these lines we compute the DS and the  functional $\mathcal{F}$. We minimize $\mathcal{F}$ over $x$, so that the functional depends now only on $\Ps$; (3) We construct the effective Hamiltonian (\ref{eq:iou0}); (4) We determine $\Ps$ as the ground state of this Hamiltonian and calculate lines of this $\Ps$ from (\ref{eq:PL_SGA})-(\ref{eq:SL_SGA}); (5) We check, whether the newly calculated lines are different (within our precision) from the input ones [those used in step (2)]. If they are different, we use the new lines as an input in step (2). If they are the same, the procedure has converged. The resulting self-consistency loop is shown in Figure \ref{fig:3}.

The alternative approach is to solve the minimization conditions as a system of equations for the lines ($P_{\veci,\vecj}$ and $S_{\veci,\vecj}$). Then, we also begin with some starting $\Ps$ as in step (1) above, and we solve the equations (\ref{eq:PL_SGA})-(\ref{eq:SL_SGA}) for variables $P_{\veci,\vecj}$ and $S_{\veci,\vecj}$. We use the GNU Scientific Library (GSL) \cite{GSL} solver \verb"gsl_multiroot_fsolver_hybrids",
which implements the hybrids algorithm. To evaluate the r.h.s. terms of (\ref{eq:PL_SGA})-(\ref{eq:SL_SGA}) we calculate the effective parameters from (\ref{eq:minCond1}). This has to be done analytically to avoid too much precision loss, because the solver estimates the Jacobian matrix by approximate methods.

The solution procedure with a self-consistency loop works faster for simple phases (e.g. it converges in a few steps for the normal phase). However, for more complicated situations (e.g. for the phase with coexistence of superconductivity and Pomeranchuk instability) the procedure can even take 1000 iterations to converge. Moreover, for the superconducting phase analysis the above procedure goes away from the minimum. Therefore, damping factors need to be introduced to ensure convergence (analogous to the situation for the Newton method in 1 dimension). Explicitly, when going from step (5) to step (2) we take as the new input lines\footnote{The damping procedure can also be introduced for the effective parameters $t^{\rm eff}_{\veci,\vecj}$ and $\Delta^{\rm eff}_{\veci,\vecj}$.} $[P_{\veci, \vecj}^{(n+1), (2)}]$ a mixture of the previous input lines $[P_{\veci, \vecj}^{(n), (2)}]$ and those newly calculated $[P_{\veci, \vecj}^{(n), (5)}]$ in step (5). Explicitly, we use
\eq
P_{\veci, \vecj}^{(n+1), (2)} = (1-\lambda) [ P_{\veci, \vecj}^{(n), (2)} ] + \lambda [ P_{\veci, \vecj}^{(n), (5)} ], \label{eq:damping}
\eqx
and analogously for the superconducting lines. The first superscript in (\ref{eq:damping}) denotes the iteration number, and the second one the step of the procedure. The choice of the damping factor $\lambda \in (0,1]$ is not trivial, as for low values the convergence is very slow, whereas for too high values, the procedure may not converge at all. We choose $\lambda$ based on the change of the lines in the last two steps.

The scheme that uses a solver instead of the self-consistency loop does not suffer from the above problem. Hovewer, it requires the computation of the Jacobian matrix elements at some (not all) steps of the procedure, for which a calculation of the derivatives of the equations with respect to all the variables (i.e. $\Ps$ lines) is necessary. If the number of these variables is of the order of 30, as can be the case for the superconducting phase, then computing the Jacobian matrix costs the same amount of time as 30 iterations of the self-consistency loop. Still, for complicated phases the solver-based procedure converges faster than that with the self-consistency loop.

The scheme without the self-consistency loop has yet another important advantage. Namely, it can be naturally supplied with the condition for the fixed number of particles. This is achieved by solving together with (\ref{eq:PL_SGA})-(\ref{eq:SL_SGA}), the equation $n_G(\Ps) - n_{\rm fixed} = 0$ for the variable $\mu_G$. The reason for using a method of solution that works with a constant number of particles $n_G$ can be the appearance of a phase separation in the studied regime. Then, a procedure working with constant $\mu_G$ can fail to converge in the region with phase separation \cite{DEtJ}. Another situation is when the system properties are to be studied as a function of parameters other than the doping $\delta = 1-2 n_G$ (e.g. as a function of the Hubbard $U$) or when two phases are to be compared at exactly the same number of particles (e.g. to compute the condensation energy or to compare their Fermi surfaces).

\section{Summary}

In this paper two schemes of minimizing the ground state energy (or grand potential) for the Gutzwiller wave function have been presented. They have been shown to be equivalent on the example of the superconducting phase analysis. While the final result of both of them is the same, they differ in technical aspects and the difficulty to obtain this result. Using both of them can serve as a simple cross-check of the obtained minimization conditions. We also discussed two ways of solving the minimization conditions and how to incorporate the condition for a fixed number of particles into one of them.

\section*{Acknowledgements}

I would like to thank J. Spa{\l}ek, J. B\"{u}nemann, and M. Wysoki\'{n}ski for discussions and comments on the manuscript. The work was supported by the Foundation for Polish Science (FNP) under the `TEAM' program. I also acknowledge the hospitality of the Leibniz Universit\"{a}t in Hannover where a large part of the work was performed.


\end{document}